\documentstyle[12pt,amsmath,amsfonts,amssymb]{article}
\renewcommand{\div}{\mathop{\rm div}\nolimits}

\makeatletter
\renewcommand{\section}{\@startsection{section}{1}{0pt}%
{3.5ex plus 1ex minus .2ex}{2.3ex plus .2ex}%
{\large\bf}}
\makeatother
\author{Alexey V. Golovnev  \\
\quad {\small alex@amber.ff.phys.spbu.ru}\\
{\small \it Saint-Petersburg State University, Saint-Petersburg, Russia}}
\title{THIN LAYER QUANTIZATION\\ IN HIGHER DIMENSIONS\\ AND CODIMENSIONS}
\date{ }
\begin{document}
\maketitle
\abstract

We consider the thin layer quantization with use of only the most elementary notions of differential geometry.
We consider this method in higher dimensions and get an explicit formula for quantum potential. 
For codimension 1 surfaces the quantum potential is presented in terms of principal curvatures,
and equivalence with Prokhorov quantization method is proved. It is shown that, in contrast with
original da Costa method, Prokhorov quantization can be generalized directly to higher
codimensions.

\newpage
\section{Introduction}
We consider free particle motion on a curved surface. Probably, the first quantum theory of it was
proposed by Podolsky in [1]; he postulated that the Hamiltonian operator  is
${\hat H}=-\frac{\hbar^2}{2}\Delta_{LB}$ with $\Delta_{LB}$ being the Laplace-Beltrami operator
on the surface. One may try to get the Podolsky theory by some quantization procedure.
Different methods of quantization yield various results which usually have the following general form:
\begin{equation}
\label{general}
{\hat H}=-\frac{\hbar^2}{2}\Delta_{LB}+V_{q}(x),
\end{equation}
the function $V_{q}(x)$ is commonly called "quantum potential".
The Dirac canonical quantization [2] and the abelian
conversion method [3,4] were discussed in our previous work [5] for surfaces of 
codimension 1 \footnote{The codimension is the difference between the bulk space dimension
and the dimension of the submanifold. In particular, codimension 1 surfaces can be defined by
one equation and have one-dimensional normal space at each point. For these surfaces all notions of
classical differential geometry are valid after obvious generalizations in the number of coordinates.
For example, in the next section we would use the principal curvatures [6] of codimension 1 surfaces.
At the same time, higher-codimensional surfaces present certain difficulties, and we encounter
some of them in the section 6.} in ${\mathbb R}^n$
This article is devoted to another theory: the thin layer quantization. In this approach the particle moves
between two equidistant infinite potential walls [7] or it is subject to some potential force
which in a proper limit makes it moving strictly along the surface [8].

The thin layer method seems to be natural for description of low dimensional motions in
nanoelectronics. Recent progress in nanotechnology caused a great activity in the field.
Free particle energy spectrum was investigated for thin layers around cylinders [9], tori [10]
and arbitrary surfaces of revolution [11]. In this article we clarify the general properties
of the thin layer method and establish its equivalence with Prokhorov quantization
procedure [12,13] for codimension 1 surfaces. For a vast majority of higher codimensional cases
the thin layer method fails to yield meaningfull results [14], at least of the general form
(\ref{general}), while the Prokhorov quantization can be generalized directly
as we show in section 6.

In section 2 we describe the thin layer quantization method in a geometrically clear manner
 which allows us to deal with any codimension 1 surface in ${\mathbb R}^n$ without any complications in comparison
 with 2-dimensional surfaces in  ${\mathbb R}^3$. In contrast with [8] we use only the most elementary notions
 of differential geometry. In section 3 we establish the equivalence of the thin layer approach with Prokhorov quantization method
 [12,13]. In section 4 a method of quantization with Hermitian momenta recently proposed by Encinosa [15] is analysed.
In sections 5 and 6 we consider surfaces of codimension greater than 1.
 
 \section {Quantization for codimension 1 surfaces}
 We consider $(n-1)$-dimensional smooth surface in ${\mathbb R}^n$ and two infinite potential walls
 at the distance $\delta\to 0$ from the surface. Free quantum particle moves in the thin layer of width
 $2\delta$ between these potential walls. We introduce a curvilinear coordinate system in which
 $|x_n|$ equals the distance from the surface to the given point, and the coordinate lines
 of $x_1,\ldots,x_{n-1}$ are orthogonal to coordinate lines of $x_n$. We have the boundary condition
 $\left.\Psi\right|_{x_n=\delta}=\left.\Psi\right|_{x_n=-\delta}=0$ and Hamiltonian 
 ${\tilde H}=-\frac{\hbar^2}{2}{\tilde\Delta}$ with $\tilde\Delta$ being the Laplace
 operator in ${\mathbb R}^n$,
 $${\tilde\Delta}=\sum\limits_{i=1}^n\sum\limits_{k=1}^n{\tilde g}^{-1/2}\partial_i{\tilde g}^{1/2}{\tilde g}^{ik}\partial_k
={\partial_n}^2+\left({\tilde g}^{-1/2}\partial_n{\tilde g}^{1/2}\right)\partial_n+\Delta_{LB},$$
$${\tilde g}_{ik}=\left (
\begin{matrix}
g_{ab}&0 \\ 
0&1
\end{matrix}
\right)$$
where $\Delta_{LB}$ is the Laplace-Beltrami operator on the surface $x_n=const$. One can prove
[5] that ${\tilde g}^{-1/2}\partial_n{\tilde g}^{1/2}=\div\overrightarrow n$ with
$\overrightarrow n$ being a unit normal vector to the surface, hence
\begin{equation}
\label{Laplace}
{\tilde\Delta}={\partial_n}^2+\div(\overrightarrow n)\cdot\partial_n+\Delta_{LB}.
\end{equation}
Indeed, let us consider two surfaces, $x_n=0$ and $x_n=\epsilon$.
Suppose we have an infinitesimal
area $dS$ at the surface $x_n=0$.
We denote the corresponding area element on $x_n=\epsilon$ surface by $dS^{\prime}$:

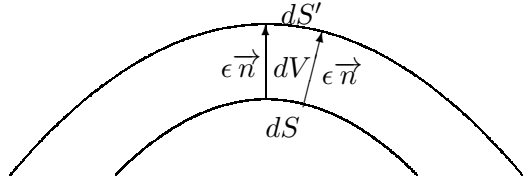
\begin{picture}(100,30)
\qbezier(30,0)(50,20)(70,0)
\qbezier(16,0)(50,40)(84,0)
\put(50,10){\vector(0,1){10}}
\put(55,9){\vector(1,4){2.5}}
\put(50,5){$dS$}
\put(52,20){$dS^{\prime}$}
\put(51,13){$dV$}
\put(44,13){$\epsilon\overrightarrow n$}
\put(57.5,12){$\epsilon\overrightarrow n$}
\end{picture}

\vspace{4ex}

\noindent We have $\div({\overrightarrow n})=\frac{dS^{\prime}-dS}{dV}+{\cal O}(\epsilon)=\frac{dS^{\prime}-dS}{\epsilon dS}+{\cal O}(\epsilon)$,
so ${\tilde g}^{-1/2}\partial_n{\tilde g}^{1/2}=\div(\overrightarrow n)$.

Now, let's take some point at the surface and consider another coordinate system
$y_1,\ldots,y_n$. We choose it to be Cartesian and such that at the given point
the tangent paraboloid of the surface is presented in its canonical form:
$y_n=\frac{1}{2}\sum\limits_{a=1}^{n-1}k_ay_a^2$. So, in the vicinity of
the chosen point ($\overrightarrow y=0$) the equation of surface is
$y_n=\frac{1}{2}\sum\limits_{a=1}^{n-1}k_ay_a^2+{\cal O}(y_a^3)$,
$k_a$-s are the principal curvatures. The unit normal is
$n_a=\frac{k_ay_a}{\sqrt{1+\sum\limits_{a=1}^{n-1}k_a^2y_a^2}}+
{\cal O}(y_a^2)=k_ay_a+{\cal O}(y_a^2)$, $n_n=-1+{\cal O}(y_a^2)$ and
\begin{equation}
\label{div}
\div{\overrightarrow n}=\sum_{a=1}^{n-1}k_a+{\cal O}(y_a).
\end{equation}

The surface $x_n=\epsilon$ can be obtained by
${\overrightarrow y}\longrightarrow{\overrightarrow y}^{\prime}=
{\overrightarrow y}+\epsilon\overrightarrow n$, and
$dy_a^{\prime}=dy_a(1+\epsilon k_a+{\cal O}(y_a))$. 
It yields $\frac{dS^{\prime}}{dS}=\frac{\prod\limits_{a=1}^{n-1}(1+{\cal O}(y_a^{\prime 2}))dy_a^{\prime}}
{\prod\limits_{a=1}^{n-1}(1+{\cal O}(y_a^{2}))dy_a}=\prod\limits_{a=1}^{n-1}
(1+\epsilon k_a)+{\cal O}(y_a)$ near the point $\overrightarrow y=0$. At the line
$y_a=0 \quad\forall a=1,\ldots,n-1$ one has
\begin{equation}
\label{areas}
\frac{dS^{\prime}}{dS}=1+\epsilon\sum_{a=1}^{n-1}k_a+\frac{1}{2}\epsilon^2\left(
\left(\sum_{a=1}^{n-1}k_a\right)^2-\sum_{a=1}^{n-1}k_a^2\right)+{\cal O}(\epsilon^3).
\end{equation}
The relation (\ref{areas}) is valid at every point of the surface provided
that one takes the principal curvatures at the same point.

Following  [7,8] we introduce a new wave function $$\chi (x)=\Psi (x)\sqrt{\frac{dS^{\prime}}{dS}}.$$
It is natural because $$\int\limits_{|x_n|\leq\delta}dV|\Psi (x)|^2=\int\limits_{-\delta}^{\delta}dx_n
\int dS|\chi (x)|^2,$$ 
so that the normal and tangential coordinates are completely separated. For the lowest
energy solutions the nornal motion is restricted only to the factor of $\cos\frac{\pi x_n}{2\delta}$,
and the integration over $x_n$ yields just the constant number. It means that the conservation
of norm for $\chi(x)$ is satisfied.
From (\ref{Laplace})-(\ref{areas}) one gets
\begin{multline*}
{\tilde\Delta}\Psi (x)={\tilde\Delta}\frac{\chi (x)}{\sqrt{\frac{dS^{\prime}}{dS}}}=
{\Delta_{LB}}\frac{\chi (x)}{\sqrt{\frac{dS^{\prime}}{dS}}}+
\frac{\partial_n^2\chi (x)}{\sqrt{\frac{dS^{\prime}}{dS}}}+\chi(x)\partial_n^2
\frac{1}{\sqrt{\frac{dS^{\prime}}{dS}}}+\\+2\partial_n\chi(x)
\partial_n\frac{1}{\sqrt{\frac{dS^{\prime}}{dS}}}+
\div\overrightarrow n\cdot\frac{\partial_n\chi(x)}{\sqrt{\frac{dS^{\prime}}{dS}}}+
\div\overrightarrow n\cdot\chi(x)\partial_n\frac{1}{\sqrt{\frac{dS^{\prime}}{dS}}}=\\=
{\Delta}_{LB}\chi (x)+\partial_n^2\chi(x)+
\left(\frac{1}{2}\sum_{a=1}^{n-1}k_a^2-\frac{1}{4}\left(\sum_{a=1}^{n-1}k_a\right)^2\right)\chi(x)
+{\cal O}(x_n).
\end{multline*}
For the lowest energy levels we have
\begin{equation}
\label{factor}
\chi(x_1,\ldots,x_n)=f(x_1,\ldots,x_{n-1})\cos\frac{\pi x_n}{2\delta}
\end{equation}
and,
after taking $\delta\to 0$ limit and subtracting an infinite (proportional to
$1/{\delta^2}$) energy, the Hamiltonian
\begin{equation}
\label{Hamilton}
{\hat H}=-\frac{\hbar^2}{2}\Delta_{LB}+\frac{\hbar^2}{8}
\left(\left(\sum_{a=1}^{n-1}k_a\right)^2-2\sum_{a=1}^{n-1}k_a^2\right)
\end{equation}
is obtained. In simple cases the factorization (\ref{factor}) works very good
even for not so small values of $\delta$ [11].

What was presented before (infinite potential walls) is rather the approach of [7] than of [8].
But the difference is not very important. One could use an appropriate
confining potential instead of infinite walls. It would lead to 
the lowest energy level function of the potential
$V_{conf}(\frac{x_n}{\delta})$ instead of $\cos\frac{\pi x_n}{2\delta}$ and to
another infinite energy.

Hamiltonian (\ref{Hamilton}) contains quantum potential
$$V_q=\frac{\hbar^2}{8} 
\left(\left(\sum_{a=1}^{n-1}k_a\right)^2-2\sum_{a=1}^{n-1}k_a^2\right).$$
For two-dimensional surfaces in ${\mathbb R}^3$ we get the result of da Costa [8]:
$V_q=-\frac{\hbar^2}{8}(k_1-k_2)^2$. For a sphere $k_a=\frac{1}{R}$ and the potential
is $V_q=\frac{\hbar^2(n-1)(n-3)}{8R^2}$.

Physically this quantization can describe lower dimensional motions
in nanoelectronics provided that the restricting potential makes a
layer of uniform effective width. Of course, it would be still a severe problem
to check this potential experimentally due to the great energy\footnote{Actually, it is not so great in comparison with the kinetic energy
of chaotic motion at ordinary temperatures. It means that the method would not work
well due to excitation of higher energy levels of the transverse motion.
So, a cryogenic experiment is needed.}
proportional to
the width of layer powered by $-2$. 
This energy is constant on the physical surface 
if the potential depends only on a distance from the surface. For more complex
potentials the results may differ. One can get different infinite energies in
different parts of the surface. It would mean infinite tangential forces, of course.
But if variation of width becomes smaller and smaller while $\delta\to 0$ it's obviously possible
to obtain some finite additional potential.

\section{Equivalence with Prokhorov quantization}
We should mention that there is one more method of quantization proposed
by Prokhorov [12]. The motion of a particle is considered as a system with two
second class constraints but the only one condition is imposed on the physical sector:
${\hat P}_n\Psi_{phys} (x)=0$ with ${\hat P}_n=-i\hbar
\frac{1}{{\tilde g}^{1/4}}\frac{\partial}{\partial x_n}{\tilde g}^{1/4}$. It means
that
\begin{equation}
\label{Prokhorov}
\partial_n\left(\sqrt{\frac{dS^{\prime}}{dS}}\Psi_{phys}(x)\right)=0.
\end{equation}
Having solved some task by this method, one should put $x_n=0$ in the results
{\it after} all the differentiations over $x_n$ are performed. 
Due to (\ref{Prokhorov}) the probability to find a particle at the distance $|x_n|$ from
the surface {\it does not depend} on the value of $x_n$, and we choose one value
we need. For Prokhorov's view see [12,13].

From (\ref{Prokhorov}) and (\ref{areas}) we conclude that
\begin{multline*}
\partial_n\Psi_{phys}(x)=-\frac{\Psi_{phys}(x)}{\sqrt{\frac{dS^{\prime}}{dS}}}
\partial_n\sqrt{\frac{dS^{\prime}}{dS}}=\\=
-\Psi_{phys}(x)\left(\frac{1}{2}\sum_{a=1}^{n-1}k_a-\frac{1}{2}x_n\sum_{a=1}^{n-1}k_a^2
+{\cal O}(x_n^2)\right),
\end{multline*}
\begin{equation*}
\partial_n^2\Psi_{phys}(x)=\Psi_{phys}(x)\left(\frac{1}{4}\left(\sum_{a=1}^{n-1}k_a\right)^2+
\frac{1}{2}\sum_{a=1}^{n-1}k_a^2+{\cal O}(x_n)\right).
\end{equation*}
Now (\ref{Laplace}) and (\ref{div}) yield:
\begin{multline*}
-\frac{\hbar^2}{2}{\tilde\Delta}\Psi_{phys}(x)=
-\frac{\hbar^2}{2}\Delta_{LB}\Psi_{phys}(x)+\\+\frac{\hbar^2}{8}
\left(\left(\sum_{a=1}^{n-1}k_a\right)^2-2\sum_{a=1}^{n-1}k_a^2\right)\Psi_{phys}(x).
\end{multline*}
So, the quantum potential coincides with the one obtained by the thin layer method.
The reason is clear. The lowest energy level wave functions (in the model with
two infinite potential walls) have nodes at $x_n=\pm\delta$ and the bunch at
$x_n=0$: $\partial_n\chi=0$ or, equivalently, ${\hat P}_n\Psi=0$. The methods of da Costa and Prokhorov are equivalent
(disregarding the infinite energy of the thin layer quantization).

Actually, one could use the calculation of ${\tilde\Delta}\Psi (x)$ from section 2 and
put $\partial_n\chi\equiv 0$ (but note that in Prokhorov method one uses $\Psi$ as a
wave function, not $\chi$).

\section{On Hermitian momenta of Encinosa}
Recently Encinosa [15] proposed one more quantization method for a constrained free motion.
The starting point is the Hamiltonian in curvilinear coordinates $x_i$, \ 
$H=\frac12\left(\sum\limits_{i-1}^{n-1}\frac{p_i^2}{h_i(x)}+p_n^2\right)$, and the recipe
is simple: $p_i\longrightarrow{\hat p_i}=-i\hbar{\tilde g}^{-1/4}\partial_i{\tilde g}^{1/4}$
followed by the thin layer method. According to these rules we have in our notations
\begin{multline*}
{\hat H}\frac{f(x_1,\ldots,x_{n-1})\cos\frac{\pi x_n}{2\delta}}{\sqrt{\frac{dS^{\prime}}{dS}}}=
\frac12\sum\limits_{i-1}^{n-1}\frac{\hat p_i^2}{h_i(x)}\frac{f(x_1,\ldots,x_{n-1})\cos\frac{\pi x_n}{2\delta}}{\sqrt{\frac{dS^{\prime}}{dS}}}-\\
-\frac12\frac{\hbar^2f(x_1,\ldots,x_{n-1})}{\sqrt{\frac{dS^{\prime}}{dS}}}\ \partial_n^2\cos\frac{\pi x_n}{2\delta}.
\end{multline*}

Effectively this result may be considered as zero quantum potential 
(geometric potential in the terminology of [15]).
We should make two
important comments on this point:

a) Quantization in curvilinear coordinates is dangerous, because the results usually
depend on the choice of coordinate system. Nevertheless, the meaning of the method considered 
could be clear if momenta operators were self-adjoint, but it's not the case even for 
spherical coordinates on the sphere.

b) Strictly speaking, the recipe is not defined correctly, because the operator odering problem in
$\frac{\hat p_i^2}{h_i(x)}$ terms is not solved. It is not difficult to deduce
the correct ordering for the zero potential theory from the relation
$-\hbar^2{\tilde\Delta}=-\hbar^2\sum\limits_{i=1}^n\sum\limits_{k=1}^n{\tilde g}^{-1/2}\partial_i{\tilde g}^{1/2}{\tilde g}^{ik}\partial_k=
\sum\limits_{i=1}^n{\tilde g}^{-1/4}{\hat p_i}{\tilde g}^{1/4}{\tilde g}^{ii}{\hat p_i}$
in any orthogonal coordinate system. But this particular ordering is not natural {\it a priori}.
Let's turn to the case of ${\mathcal S}^2$ with stereographic coordinates
$x_1=2(R+x_3)\cot\left (\frac{\vartheta}{2}\right )\cos\varphi$ and 
$x_2=2(R+x_3)\cot\left (\frac{\vartheta}{2}\right )\sin\varphi$.
In [16] it was shown that in these coordinates the Laplace-Beltrami operator
on the sphere of radius $R+x_3$ equals
$$-\frac{{\hbar}^2}{2}\Delta_{LB}=\left (1+\frac{x_1^2+x_2^2}%
{4(R+x_3)^2}\right )\cdot%
\frac{\left (\hat p_1^2+%
\hat p_2^2\right )}{2}\cdot\left (1+\frac{x_1^2+%
x_2^2}%
{4(R+x_3)^2}\right ).$$
Even less natural  it would seem  for ${\mathcal S}^n$ in ${\mathbb R}^{n+1}$:
\begin{multline*}
-\frac{{\hbar}^2}{2}\Delta_{LB}=\frac12{\left (1+\frac{x_1^2+\ldots+%
x_n^2}{4(R+x_{n+1})^2}%
\right )}^{n/2}
\cdot\\   \cdot
\sum_{i=1}^{n}\left (\hat p_i {\left (1+\frac{x_1^2+%
\ldots+x_n^2}{4(R+x_{n+1})^2}%
\right )}^{2-n}\hat p_i\right )\cdot%
 {\left (1+\frac{x_1^2+\ldots+x_n^2}{4(R+x_{n+1})^2}\right )}^{n/2}.  
\end{multline*}
And, of course, for coordinates of section 5 in [16] the situation would not be better.

\section{Elementary cases of codimension $> 1$}
In general let's consider $m$-dimensional smooth surface in ${\mathbb R}^n$ represented by its tangent
paraboloid at some point:
\begin{equation}
\label{codim}
y_{\alpha}=\frac{1}{2}\sum_{a=1}^m\sum_{b=1}^mk^{(\alpha)}_{ab}y_ay_b+{\cal O}(y_a^3),
\end{equation}
$\alpha=m+1,\ldots,n$, $k^{(\alpha)}_{ab}=k^{(\alpha)}_{ba}$. In general $n-m$ curvature forms
$k^{(\alpha)}$ can't be diagonalized simultaneously and the notion of principal curvatures
does not exist, but in this section we treat the simplest cases for which the diagonalization
can be performed.

First of all, for $1$-dimensional manifolds (curves) curvature forms $k^{(\alpha)}$ are just real
numbers. In this case by a rotation in the space of $y_{\alpha}$ one can get
$y_2=\frac{1}{2}ky_1^2+{\cal O}(y_1^3)$, $y_3,\ldots,y_n={\cal O}(y_1^3)$. The unit normal vectors
are $n_1^{(2)}=ky_1+{\cal O}(y_1^2)$, $n_2^{(2)}=-1
+{\cal O}(y_1^2)$, $n_3^{(2)}=\ldots=n_n^{(2)}={\cal O}(y_1^2)$; $n_i^{(\alpha)}=-\delta_{i\alpha}+
{\cal O}(y_1^2)$ for $\alpha\geq 3$. We have ${\overrightarrow n}^{(\alpha)}{\overrightarrow n}^{(\beta)}=\delta_{\alpha\beta}+{\cal O}(y_1^2)$
and after the transformation ${\overrightarrow y}\to{\overrightarrow y}^{\prime}={\overrightarrow y}+
\sum\limits_{\alpha=2}^n\epsilon_{\alpha}{\overrightarrow n}^{(\alpha)}$ one gets
$dy_1^{\prime}=(1+\epsilon_2k+{\cal O}(y_1))dy_1$, $dy_{\alpha}^{\prime}=(1+{\cal O}(y_1))dy_{\alpha}$.

Now we consider a smooth family of such coordinate systems and normals along the curve.
We introduce a new curvilinear coordinate system in which $x_1$ is just the length along the curve
while $x_{\alpha}={\overrightarrow n}^{(\alpha)}\cdot{\overrightarrow r}$ where $\overrightarrow r$ is the minimal
norm radius vector from the curve to a given point and ${\overrightarrow n}^{(\alpha)}$ is taken
at the same point on the curve as $\overrightarrow r$. In this coordinate system we have
$${\tilde g}_{ik}=\left (
\begin{matrix}
(1+x_2k)^2&0 \\ 
0&I
\end{matrix}
\right)$$ and
${\tilde\Delta}=\Delta_{c}+\Delta_{n}+\left(\frac{1}{1+x_2k}\partial_2(1+x_2k)\right)\partial_2=
\Delta_{c}+\Delta_{n}+\frac{k}{1+x_2k}\partial_2$ where $\Delta_{c}$ is Laplace-Beltrami operator
on a curve $x_{\alpha}=const$ and $\Delta_{n}=\sum\limits_{\alpha=2}^n\partial_{\alpha}^2$ is Laplace
operator  in a hyperplane $x_1=const$.

To proceed with the thin layer quantization we introduce a thin layer $\sum\limits_{\alpha=2}^nx_{\alpha}^2=\delta^2$
around the curve and a wave function $\chi(x)=\sqrt{1+x_2k}\ \Psi(x)$ such that
$\int dx_1\int dS_{normal}|\chi(x)|^2=\int dV|\Psi(x)|^2$:
\begin{multline*}
{\tilde\Delta}\Psi(x)={\tilde\Delta}\frac{\chi(x)}{\sqrt{1+x_2k}}=
{\Delta}_c\chi(x)+{\Delta}_n\chi(x)+\frac{k^2}{4}\chi(x)+{\cal O}(x_{\alpha}).
\end{multline*}
After subtracting an infinite energy due to ${\Delta}_n\chi(x)$ it yields the quantum potential
$V_q=-\frac{\hbar^2}{8}k^2$ as in [8].

We should mention that it is very important for this result that the thin layer is spherical at every point 
of the curve. Indeed, suppose we have a straight line in ${\mathbb R}^3$. We can first embed  a cylinder 
of any radius $R$ into ${\mathbb R}^3$ and after that, with much more thin layer, we embed a line into the 
cylinder. The first embedding results in $V_q=-\frac{\hbar^2}{8R^2}$ and the second one changes nothing
because the line is geodesic in the cylinder. Note that it is possible to generalize our considerations
to the case of embedding into non-Euclidean spaces in a sense that, if one has a free particle 
Hamiltonian (probably, with some quantum potential) in a Riemannian manifold, he may consider a thin
layer of constant width around any codimension 1 submanifold. For a cylinder it would literally reproduce
the demonstrations of the section 2 because in intrinsic geometry the cylinder is flat.

As another simple example we consider $2$-dimensional flat torus isometrically embedded in ${\mathbb R}^4$:
\begin{equation*}
\left\{
\begin{aligned}
x_1^2+x_2^2=R_1^2,\\
x_3^2+x_4^2=R_2^2.
\end{aligned}
\right.
\end{equation*}
We use the following coordinate system:
\begin{equation*}
\left\{
\begin{aligned}
\phi_1=\arctan\frac{x_2}{x_1},\\
\phi_2=\arctan\frac{x_4}{x_3},\\
r_1=\sqrt{x_1^2+x_2^2}-R_1,\\ 
r_2=\sqrt{x_3^2+x_4^2}-R_2
\end{aligned}
\right.
\end{equation*}
with
$${\tilde g}_{ik}=\left (
\begin{matrix}
(r_1+R_1)^2&0&0&0 \\ 
0&(r_1+R_1)^2&0&0\\
0&0&1&0\\
0&0&0&1
\end{matrix}
\right)$$
and ${\tilde\Delta}=\Delta_t+\Delta_n+\frac{1}{r_1+R_1}\partial_{r_1}+\frac{1}{r_2+R_2}\partial_{r_2}$ 
where $\Delta_t$ is the Laplace-Beltrami operator on the torus $r_1=const$, $r_2=const$.
We have the volume element $dV=(r_1+R_1)(r_2+R_2)dr_1dr_2d\phi_1d\phi_2$, the wave function
$\chi=\sqrt{(1+\frac{r_1}{R_1})(1+\frac{r_2}{R_2})}\ \Psi$ and the quantum potential
$V_q=-\frac{\hbar^2}{8}(k_1^2+k_2^2)$ with $k_a=\frac{1}{R_a}$.

\section{General theory}
In general the geometric construction is very similar to that of $1$-dimensional case. The unit normals to the surface
(\ref{codim}) at $\overrightarrow y=0$ are 
$n_{\beta}^{(\alpha)}=\delta_{\alpha\beta}(-1+{\cal O}(y_a^2))$, 
$n_a^{(\alpha)}=\sum\limits_{b=1}^mk^{(\alpha)}_{ab}y_b+{\cal O}(y_a^2)$,
$\alpha,\beta=m+1,\ldots,n$,\quad $a,b=1,\ldots,m$,
 and after the replacement
${\overrightarrow y}\to{\overrightarrow y}^{\prime}={\overrightarrow y}+
\sum\limits_{\alpha=m+1}^n\epsilon_{\alpha}{\overrightarrow n}^{(\alpha)}$
one gets $$dy_a^{\prime}=\sum\limits_{b=1}^m\left(\delta_{ab}+\sum\limits_{\alpha=m+1}^{n}\epsilon_{\alpha}
k^{(\alpha)}_{ab}+{\cal O}(y_c)\right)dy_b,$$ $dy_{\alpha}^{\prime}=(1+{\cal O}(y_c))dy_{\alpha}$.
It means that in curvilinear coordinate system
\begin{multline*}
g_{ab}(x_1,\ldots,x_n)=\sum_{c=1}^m
\left(\delta_{ca}+\sum\limits_{\alpha=m+1}^{n}x_{\alpha}
k^{(\alpha)}_{ca}\right)\cdot\\ \cdot
\left(\delta_{cb}+\sum\limits_{\beta=m+1}^{n}x_{\beta}
k^{(\beta)}_{cb}\right)f_{ab}(x_1,\ldots,x_m)
\end{multline*}
with $x_{\alpha}={\overrightarrow n}^{(\alpha)}\cdot{\overrightarrow r}$ as in the previous section
and $f_{ab}(x_1,\ldots,x_m)=g_{ab}(x_1,\ldots,x_m,0,\ldots,0)$. In this system if some point
has coordinates $(x_1,\ldots,x_n)$ then its position can be found by adding the vector 
$\sum\limits_{\alpha=m+1}^nx_{\alpha}{\overrightarrow n}^{(\alpha)}$ to the radius-vector of the initial 
surface point with
coordinates $(x_1,\ldots,x_m)$. If we choose $x_a=y_a$ at the surface $x_{\alpha}=0$, then at the $x_a=0$ hyperplane
$f_{ab}=\delta_{ab}$,  $x_{\alpha}=\epsilon_{\alpha}$ and
$g_{ab}=\delta_{ab}+2\sum\limits_{\alpha}\epsilon_{\alpha}k_{ab}^{(\alpha)}+
\sum\limits_{\alpha,\beta,c}\epsilon_{\alpha}\epsilon_{\beta}k_{ac}^{(\alpha)}k_{bc}^{(\beta)}$.
An easy computation yields
\begin{multline*}
g=1+2\sum\limits_{\alpha,a}\epsilon_{\alpha}k_{aa}^{(\alpha)}+
2\sum\limits_{\alpha,\beta,a,b}\epsilon_{\alpha}\epsilon_{\beta}k_{aa}^{(\alpha)}k_{bb}^{(\beta)}-
2\sum\limits_{\alpha,\beta,a}\epsilon_{\alpha}\epsilon_{\beta}k_{aa}^{(\alpha)}k_{aa}^{(\beta)}+\\+
3\sum\limits_{a,b}\left(\sum\limits_{\alpha}\epsilon_{\alpha}k_{ab}^{(\alpha)}\right)^2-
2\sum\limits_{a}\left(\sum\limits_{\alpha}\epsilon_{\alpha}k_{aa}^{(\alpha)}\right)^2+{\cal O}(\epsilon^3).
\end{multline*}

The problem is that the coordinates $x_i$ are not orthogonal. Parallel translations
of the surface along one of the normals break its orthogonality to other normals. Indeed,
we have $\frac{\partial n_b^{(\alpha)}}{\partial y_a}=k_{ab}^{(\alpha)}+{\cal O}(y_c)$ and 
$\frac{\partial n_{\beta}^{(\alpha)}}{\partial y_a}={\cal O}(y_c)$. We supose that all normals have
unit length, hence $0=\frac{\partial \overrightarrow n^{(\alpha)}}{\partial y_a}\cdot{\overrightarrow n}^{(\alpha)}=
\sum\limits_{b,c}k_{bc}^{(\alpha)}y_ck_{ab}^{(\alpha)}-\frac{\partial n_{\alpha}^{(\alpha)}}{\partial y_a}+{\cal O}(y_c^2)$ and
$\frac{\partial n_{\alpha}^{(\alpha)}}{\partial y_a}=\sum\limits_{b,c}k_{ab}^{(\alpha)}k_{bc}^{(\alpha)}y_c+{\cal O}(y_c^2)$.
Now we see that parallel translations lead to violation of orthogonality condition because
$\frac{\partial \overrightarrow n^{(\beta)}}{\partial y_a}\cdot{\overrightarrow n}^{(\alpha)}=
\sum\limits_{b,c}\left(k_{ab}^{(\beta)}-k_{ab}^{(\alpha)}\right)k_{bc}^{(\alpha)}y_c+{\cal O}(y_c^2)\neq0$.

We introduce a new notation $f_{\alpha a}^{(\beta)}=f_{a \alpha}^{(\beta)}=
\sum\limits_{b,c}\left(k_{ab}^{(\beta)}-k_{ab}^{(\alpha)}\right)k_{bc}^{(\alpha)}y_c$ and find the 
metric tensor
$${\tilde g}_{ik}=\left (
\begin{matrix}
g_{ab}&\sum\limits_{\gamma}f_{a \beta}^{(\gamma)}\epsilon_{\gamma}+{\cal O}(\epsilon^2) \\ 
\sum\limits_{\gamma}f_{\beta a}^{(\gamma)}\epsilon_{\gamma}+{\cal O}(\epsilon^2)&\delta_{\alpha\beta}
\end{matrix}
\right),$$
its determinant ${\tilde g}=g+{\cal O}(\epsilon^2)$ and the reciprocal tensor
$${\tilde g}^{ik}=\left (
\begin{matrix}
g^{ab}&-\sum\limits_{c,\gamma}g^{ac}f_{c \beta}^{(\gamma)}\epsilon_{\gamma}+{\cal O}(\epsilon^2) \\ 
-\sum\limits_{c,\gamma}f_{\alpha c}^{(\gamma)}g^{cb}\epsilon_{\gamma}+{\cal O}(\epsilon^2)&\delta_{\alpha\beta}
\end{matrix}
\right).$$
Da Costa concluded [14] that the thin layer quantization would not work well in this situation
because ${\tilde\Delta}$ contains terms with both derivatives $\partial_a$ and $\partial_{\alpha}$.

The situation is quite different for the method proposed in [12].
All new terms in ${\tilde\Delta}$  have coefficients of order ${\cal O}(\epsilon)$; the only suspicious term,
$\sum\limits_{a,\alpha}(\partial_{\alpha}g^{a\alpha})\partial_a=\left(-\sum\limits_{a,c,\alpha,\beta}
g^{ac}f_{c \alpha}^{(\beta)}\delta_{\alpha\beta}+{\cal O}(\epsilon)\right)\partial_a={\cal O}(\epsilon)\partial_a$,
is not dangerous
because $f_{c \alpha}^{(\alpha)}=0$.
In Prokhorov quantization method the condition $\partial_{\alpha}\left(g^{1/4}\Psi_{phys}\right)=0,\quad
\forall\alpha=m+1,\ldots,n$ is imposed. For a function
$\chi(x)=\left(\frac{g}{f}\right)^{1/4}\Psi(x)$ it means $\partial_{\alpha}\chi=0$, and we get
\begin{multline*}
{\tilde\Delta}\frac{\chi(x)}{\left(\frac{g}{f}\right)^{1/4}}
=\Delta_{LB}\frac{\chi(x)}{\left(\frac{g}{f}\right)^{1/4}}
+\chi(x)\Delta_{n}\frac{1}{\left(\frac{g}{f}\right)^{1/4}}+\\
+\chi(x)\sum_{\alpha=m+1}^{n}\left(\frac{1}{\sqrt{g}}\partial_{\alpha}\sqrt{g}\right)\partial_{\alpha}\frac{1}{\left(\frac{g}{f}\right)^{1/4}}+{\cal O}(x_{\alpha}),
\end{multline*}
$\Delta_n\equiv\sum\limits_{\alpha=m+1}^{n}\partial_{\alpha}^2$.
It is easy to see that $\frac{1}{\sqrt{g}}\partial_{\alpha}\sqrt{g}=-2g^{1/4}\partial_{\alpha}g^{-1/4}$,
and  the Hamiltonian is
${\hat H}=-\frac{\hbar^2}{2}\Delta_{LB}+V_q(x)$,
$$V_q=-\frac{\hbar^2}{2}\left.\left(g^{1/4}\Delta_ng^{-1/4}-2\sum\limits_{\alpha=m+1}^{n}\left(g^{1/4}\partial_{\alpha}g^{-1/4}\right)^2\right)
\right|_{x_{\alpha}=0}$$
One more (not too difficult) calculation yields
\begin{equation}
\label{pot}
V_q=\frac{\hbar^2}{8}\sum\limits_{\alpha=m+1}^n\left(\left(\sum\limits_{a=1}^mk_{aa}^{(\alpha)}\right)^2+
6\sum_{a=1}^m\sum_{b=1}^m\left(k_{ab}^{(\alpha)}\right)^2-8\sum_{a=1}^m\left(k_{aa}^{(\alpha)}\right)^2\right).
\end{equation}

The thin layer method would not give this answer because in this method we have $\partial_{\alpha}\chi=0$
only at the original surface and generally $\partial_{\alpha}\chi\sim\frac{1}{\delta}$, so that
terms with $\partial_{\alpha}\chi$ are not negligible and factorization analogues to (\ref{factor}) is not
a good approximation of the exact thin layer solution. It's not surprising. Let us take a small
element of the surface and its $\delta$-neighbourhood, $\delta\to0$. The well-known theorem 
states that $\int dV \Delta\Psi=\int {\overrightarrow{dS}}\cdot{\overrightarrow\bigtriangledown}\Psi$. We have
$\Delta\Psi\approx\frac{\int {\overrightarrow{dS}}\cdot{\overrightarrow\bigtriangledown}\Psi}{V}$ and in
codimension 1 case the normal projection of ${\overrightarrow\bigtriangledown}\Psi$ leads to $\frac{1}{\delta^2}$
term in $\Delta\Psi$ (because $\partial_n\Psi\sim\frac{1}{\delta}$ and $V\sim\delta$). Tangential components
of ${\overrightarrow\bigtriangledown}\Psi$ result in finite values of $\Delta\Psi$  due to $dS_{\perp}\sim\delta$.
In general case $V\sim\delta^{n-m}$ and $dS\sim\delta^{n-m}$, hence the tangential components of
${\overrightarrow\bigtriangledown}\Psi$  yield finite terms in $\Delta\Psi$ again; but normal projections gain some components orthogonal
to transverse hyperplanes ($x_a=const$). The corresponding angles are of order $\delta$ but $\partial_{\alpha}\Psi
\sim\frac{1}{\delta}$. It results in finite terms in $\Delta\Psi$. So, normal components of ${\overrightarrow\bigtriangledown}\Psi$
influence the tangential dynamics. Hence, the thin layer method of [7,8,14] {\it does not} yield the result of the
general form (\ref{general}). It means that the Prokhorov method in such cases {\it does not} correspond to the motion in the uniform thin
layer but it turns to be more powerfull as an abstract quantization method for
second class constrained systems. One could find, of course, some orthogonal coordinate system in the
whole vicinity of the surface with normals depending on $x_{\alpha}$ and, may be, he would be able to
determine a geometry of the thin layer for which some factorization analogues to (\ref{factor})
would be a good approximation. Quantum potential in this approach is likely to coincide with (\ref{pot}) 
at least with a certain realization of it, but generally the layer would not have constant width and the
set-up of the quantization would not be as easy and clear as the original one.

\newpage

\vspace*{2ex}
{\large \bf \begin{center} References. \end{center}}
\vspace{2ex}
\begin{enumerate}
\item B. Podolsky, {\it Phys.Rev.} {\bf 32}, 812 (1928).
\item P.A.M. Dirac, {\it Lectures on quantum mechanics} (Yeshiva University, N.Y., 1964).
\item L.D. Faddeev, S.L. Shatashvili, {\it Phys. Lett. B}\quad {\bf 167}, 255 (1986).
\item I.A. Batalin, E.S. Fradkin, {\it Nucl. Phys. B}\quad {\bf 279}, 514 (1987).
\item A.V. Golovnev, {\it preprint} quant-ph/0508044; to appear in {\it Int. J. Geom. Meth. Mod. Phys.}
\item A. Mishchenko, A. Fomenko, {\it A course of differential geometry and topology} (Mir
Publishers, Moscow, 1988).
\item H. Jensen, H. Koppe, {\it Ann. Phys.} {\bf 63}, 586 (1971).
\item R.C.T. da Costa, {\it Phys. Rev. A} {\bf 23}, 1982 (1981).
\item M. Encinosa, B. Etemadi, {\it Phys. Rev. A} {\bf 58}, 77 (1998).
\item M. Encinosa, L. Mott, B. Etemadi, {\it preprint} quant-ph/0409141.
\item J. Gravesen, M. Willatzen, L.C. Lew Yan Voon, {\it J. Math. Phys.} {\bf 46}, 012107 (2005).
\item L.V. Prokhorov, in {\it Proc. of VI int. conf. on Path Integrals (1998, Florence)}, p. 249 (London, World Scientific, 1999).
\item A.G. Nuramatov, L.V. Prokhorov, {\it preprint} quant-ph/0507038.
\item R.C.T. da Costa, {\it Phys. Rev. A} {\bf 25}, 2893 (1982).
\item M. Encinosa, {\it preprint} quant-ph/0508104.
\item A.V. Golovnev, L.V. Prokhorov, {\it J. Phys. A} {\bf 37}, 2765 (2004);
preprint quant-ph/0306080.
\end{enumerate}
\end{document}